# Practical Implementation of a Deep Random Generator

Thibault de Valroger [(*)]

Abstract

We have introduced in former work [5] the concept of Deep Randomness and its interest to design Unconditionally Secure communication protocols. We have in particular given an example of such protocol and introduced how to design a Deep Random Generator associated to that protocol. Deep Randomness is a form of randomness in which, at each draw of random variable, not only the result is unpredictable bu also the distribution is unknown to any observer. In this article, we remind formal definition of Deep Randomness, and we expose two practical algorithmic methods to implement a Deep Random Generator within a classical computing resource. We also discuss their performances and their parameters.

**Key words.** Deep Random, Random Generator, Perfect Secrecy, Unconditional Security, Prior Probabilities, information theoretic security

## I. Introduction

**Prior probabilities theory**

Before presenting the Deep Random assumption, it is needed to introduce Prior probability theory.

The art of prior probabilities consists in assigning probabilities to a random event in a context of partial or complete uncertainty regarding the probability distribution governing that random event. The first author who has rigorously considered this question is Laplace [10], proposing the famous *principle of insufficient reason* by which, if an observer does not know the prior probability of occurrence of 2 events, he should consider them as equally likely. In other words, if a random variable $X$ can take several values $v_1, \ldots, v_n$, and if no information regarding the prior probabilities $P(X = v_i)$ is available for the observer, he should assign them $P(X = v_i) = 1/n$ in any attempt to produce inference from an experiment of $X$.

Several authors observed that this principle can lead to conclusion contradicting the common sense in certain cases where some knowledge is available to the observer but not reflected in the assignment principle.

If we denote $\Im_<$ the set of all prior information available to observer regarding the probability distribution of a certain random variable ('prior' meaning before having observed any experiment of that variable), and $\Im_>$ any public information available regarding an experiment of $X$, it is then possible to define the set of possible distributions that are compatible with the information $\mathfrak{I} \triangleq \mathfrak{I}_< \cup \mathfrak{I}_>$ regarding an experiment of $X$; we denote this set of possible distributions as:

$$D_\mathfrak{I}$$

*(*) See contact and information about the author at last page*

The goal of Prior probability theory is to provide tools enabling to make rigorous inference reasoning in a context of partial knowledge of probability distributions. A key idea for that purpose is to consider groups of transformation, applicable to the sample space of a random variable $X$, that do not change the global perception of the observer. In other words, for any transformation $\tau$ of such group, the observer has no information enabling him to privilege $\varphi_\mathfrak{I}(v) \triangleq P(X = v|\mathfrak{I})$ rather than $\varphi_\mathfrak{I} \circ \tau(v) = P(X = \tau(v)|\mathfrak{I})$ as the actual conditional distribution. This idea has been developed by Jaynes [7], in order to avoid the incoherence brought in some cases by the imprudent application of Laplace principle.

We will consider only finite groups of transformation, because one manipulates only discrete and bounded objects in digital communications. We define the acceptable groups $G$ as the ones fulfilling the 2 conditions below:

($C1$)  Stability - For any distribution $\varphi_\mathfrak{I} \in D_\mathfrak{I}$, and for any transformation $\tau \in G$, then $\varphi_\mathfrak{I} \circ \tau \in D_\mathfrak{I}$

($C2$)  Convexity - Any distribution that is invariant by action of $G$ does belong to $D_\mathfrak{I}$

It can be noted that the set of distributions that are invariant by action of $G$ is exactly:

$$R_\mathfrak{I}(G) \triangleq \left\{ \frac{1}{|G|} \sum_{\tau \in G} \varphi_\mathfrak{I} \circ \tau \;|\; \forall \varphi_\mathfrak{I} \in D_\mathfrak{I} \right\}$$

The condition ($C2$) is justified by the fact that in the absence of information enabling the observer to privilege $\varphi_\mathfrak{I}$ from $\varphi_\mathfrak{I} \circ \tau$, he should choose equivalently one or the other distribution, but then of course the average distribution $\frac{1}{|G|} \sum_{\tau \in G} \varphi_\mathfrak{I} \circ \tau$ should still belong to the set $D_\mathfrak{I}$ of possible distributions knowing $\mathfrak{I}$.

The set of acceptable groups as defined above is denoted:

$$\Gamma_\mathfrak{I}$$

Let's consider some examples.

**Example 1:** we consider a 6-sides dice. We are informed that the distribution governing the probability to draw a given side is altered, but we have no information of what that distribution actually is, and we have no available information regarding an experiment. We want nevertheless to assign an a priori probability distribution for the draw of dice. In this very simple case, it seems quite reasonable to assign an a priori probability of 1/6 to each side. A more rigorous argument to justify this decision, based on the above, is the following: let's consider $G$ the finite group of transformation that let the dice unchanged, this group $G$ is well known, it is generated by the 3 axis 90° rotations, and has 24 elements. It is clear here that $G \in \Gamma_\mathfrak{I}$. It is also clear that, by considering a given distribution $(p_1, \dots, p_6)$, the a priori information available to the observer gives him no ground to privilege $(p_1, \dots, p_6)$ rather than $(p_{g(1)}, \dots, p_{g(6)})$ for any $g \in G$, and therefore the distribution should be of the form:

$$\left\{\left(\frac{1}{|G|}\sum_{g\in G}p_{g(i)}\right)_{i\in\{1,\ldots,6\}}\right\}_{\{(p_1,\ldots,p_6)\}}$$

It is easy to calculate that whatever is $(p_1, \ldots, p_6)$,

$$\frac{1}{|G|}\sum_{g\in G}p_{g(i)} = \frac{1}{6}\sum_{i=1}^{6}p_i = \frac{1}{6}$$

and therefore in this trivial example, the Laplace principle applies nicely.

**Example 2**: let's now suppose that we have the result $J$ of a draw of the dice. Then the symmetry disappears and the opponent may want to assign, as an extreme example, a probability

$$p_j = \Gamma_J(j) \triangleq \begin{cases} 1 \text{ if } j = J \\ 0 \text{ otherwise} \end{cases}$$

which does not follow Laplace principle although one could argue that the knowledge of one single draw is not incompatible with the distribution $\{p_j = 1/6\}$. We however clearly don't want to exclude that extreme choice from the set of theoretically valid assignment made by the observer. An applicable group is in this case the sub-group $H$ of $G$ that let the side $J$ invariant. It is easy to see that it is composed with the 4 rotations $(0°, 90°, 180°, 270°)$ whose axis is determined by the centers of the sides $J$ and $7 - J$.

$$\left\{\left(\frac{1}{|H|}\sum_{h\in H}p_{h(i)}\right)_{i\in\{1,\ldots,6\}}\right\}_{\{(p_1,\ldots,p_6)\}}$$
$$= \left\{p_J = u; p_{7-J} = v; p_s = w \ \forall s \in \{1,\ldots,6\}\backslash\{J, 7-J\} \ \middle| \ \begin{array}{l} u \geq 0 \\ v \geq 0 \\ w \geq 0 \\ u + v + 4w = 1 \end{array}\right\}$$

One could argue that in a given probabilistic situation, there may exist several groups of transformation in $\Gamma_\Im$, and in that case, the choice of a given such group may appear arbitrary to assign the probability distribution. Although that oddness is not a problem for our purpose, we can solve it when 2 reasonable conditions are fulfilled: (i) we only consider finite sample space (note that objects in digital communication theory are bounded and discrete), and (ii) we assume that $D_\Im$ is convex (which can be ensured by design of the Deep Random source). Under those 2 conditions, all groups of transformation applying on the sample space are sub-groups of the finite (large) group of permutations $\mathfrak{S}$ applying on all the possible states, and if 2 groups of transformations $G$ and $G'$ applying on the sample space are in $\Gamma_\Im$, it is easy to see that $G \vee G'$ is still a finite sub-group of $\mathfrak{S}$, it contains $G$ and $G'$ and it is still in $\Gamma_\Im$. Consequently, for any observer having the same knowledge $\Im$ of the distribution, there exists a unique maximal sub-group $T_\Im \in \Gamma_\Im$, and this one should be ideally applied to obtain a maximal restriction of the set of distribution (because it is easy to check that if $G' \subset G$, then $R_\Im(G) \subset R_\Im(G')$). We can remark also that, if the uniform distribution is in $\Omega_\Im$, then $\bigcap_{G\in\Gamma_\Im}R_\Im(G) \neq \emptyset$ because $R_\Im(G)$ then always contains the uniform distribution.

However, it is not necessary for our purpose to be able to determine such unique maximal sub-group $T_\Im$ if it exists.

Another objection could be that the definition of $D_\Im$ may appear too 'black or white' (a distribution belongs or does not belong to $D_\Im$), although real situations may not be so contrasted. This objection is not applicable to our purpose because in the case of Deep Random Generation, that we will introduce in section IV, we can ensure by design that distributions do belong to a specific set $D_\Im$. But otherwise, generally speaking, it is somehow preferable to draft the condition $(C1)$ as:

($C1$) For any distribution $\Phi$, and for any transformation $\tau$ in such group, the observer has no information enabling him to privilege $\Phi$ rather than $\Phi \circ \tau$ as the actual distribution.

**Deep Random assumption**

We can now introduce and rigorously define the Deep Random assumption. The Deep Random assumption imposes a rule enabling to make rigorous inference reasoning about an observer $\xi$ in a context where that observer $\xi$ only has partial knowledge of a probability distribution.

Let $X$ be a private random variable with values in a metric set $F$. Keeping the notations of the previous section, we still denote $\Im_<$ the set of all prior information available to observer regarding the probability distribution of $X$ ('prior' meaning before having observed any experiment of that variable), and $\Im_>$ any public information available regarding an experiment of $X$. Typically, $X$ represents a secret information that $\xi$ has to determine with the highest possible accuracy, and $\Im \triangleq \Im_< \cup \Im_>$ represents the public information available to $\xi$ (like for instance the information about the design of the Deep Random source, plus the information exchanged during a communication protocol as introduced in the following sections).

For any group $G$ of transformations applying on the sample space $F$, we denote by $\Omega_\Im(G)$ the set of all possible conditional expectations when the distribution of $X$ courses $R_\Im(G)$. In other words:

$$\Omega_\Im(G) \triangleq \{Z(\Im) \triangleq E[X|\Im] | \forall \varphi_\Im \in R_\Im(G)\}$$

Or also:

$$\Omega_\Im(G) = \left\{ Z(\Im) = \int_F v \varphi_\Im(v) dv \,|\, \forall \varphi_\Im \in R_\Im(G) \right\}$$

The **Deep Random assumption** prescribes that, if $G \in \Gamma_\Im$, the strategy $Z_\xi$ of the opponent observer $\xi$, in order to estimate $X$ from the public information $\Im$, should be chosen by the opponent observer $\xi$ within the restricted set of strategies:

$$\boldsymbol{Z_\xi \in \Omega_\Im(G)} \quad (\boldsymbol{A})$$

The Deep Random assumption can thus be seen as a way to restrict the possibilities of $\xi$ to choose his strategy in order estimate the private information $X$ from his knowledge of the public information $\Im$. It is a fully reasonable assumption as exposed in the previous section presenting Prior probability theory, because the assigned prior distribution should remain stable by action of a transformation that let the distribution uncertainty unchanged.

($A$) suggests of course that $Z_\xi$ should eventually be picked in $\bigcap_{G \in \Gamma_\Im} \Omega_\Im(G)$ (that equals to $\Omega_\Im(T_\Im)$ when $T_\Im$ exists), but it is enough for our purpose to find at least one group of transformation with which one can apply efficiently the Deep Random assumption to the a protocol in order to measure an advantage distilled by the legitimate partners compared to the opponent.

**Application of Deep Random generators to Secrecy Systems**

Shannon, in [1], established his famous impossibility result. Shannon defines perfect secrecy of a secrecy system, as its ability to equal the probability of the clear message $P(M)$ and the conditional probability $P(M|E)$ of the clear message knowing the encrypted message. In the case where the encryption system is using a shared secret key $K$ with a public transformation procedure to transform the clear message into the encrypted message, Shannon establishes that perfect secrecy can only be obtained if $H(K) \geq H(M)$.

It is a common belief in the cryptologic community that, in cases where the legitimate partners initially share no secret information (which we can write $H(K) = 0$), the result of Shannon thus means that it is impossible for them to exchange a perfectly secret (or almost perfectly secret) bit of information. The support for that belief is that, in the absence of key entropy, the conditional expectation $E[M|E]$, that is the best possible estimation of $M$ knowing $E$, is completely and equally known by all the parties (legitimate receiver and observer), as :

$$E[M|E] = \sum_m m \frac{P(E|m)P(m)}{\sum_{m'} P(E|m')P(m')}$$

and thus, that the legitimate receiver cannot gain any advantage over the observer when he tries to estimate the secret clear message from the public encrypted message. This reasoning however supposes that all the parties have a full knowledge of the distribution $P(M)$, enabling them to perform the above Bayesian inference to estimate $M$ from $E$.

Shannon himself warned the reader of [1] to that regard, but considered that this assumption is fairly reasonable (let's remember that computers were almost not yet existing when he wrote his article):

*« There are a number of difficult epistemological questions connected with the theory of secrecy, or in fact with any theory which involves questions of probability (particularly a priori probabilities, Bayes' theorem, etc.) when applied to a physical situation. Treated abstractly, probability theory can be put on a rigorous logical basis with the modern measure theory approach. As applied to a physical situation, however, especially when "subjective" probabilities and unrepeatable experiments are concerned, there are many questions of logical validity. For example, in the approach to secrecy made here,* **a priori probabilities of various keys and messages are assumed known by the enemy cryptographer.** *»*

The model of security that we have developed in [5], by enabling the legitimate partners to use a specific form of randomness where the a priori probabilities of the messages cannot be efficiently known by the observer, puts this observer in a situation where the above reasoning based on Bayesian inference no longer stands. We considered in [5] secrecy protocols where all the information exchanged by the legitimate partners $A$ and $B$ over the main channel are fully available to the passive observer $\xi$. $\xi$ has unlimited computation and storage power. $A$ and $B$ share initially no private information.

Beyond being capable to Generate random bit strings, Publish bit strings on the main channel, Read published bit strings from the main channel, Store bit strings, and Make computation on bit strings, the legitimate partners $A$ and $B$ are also capable to generate bit strings with Deep Randomness, by using their Deep Random Generator (DRG). A DRG is a random generator that produces an output with a probability distribution that is made unknown to an external observer.

$\xi$ is capable to Read all published bit strings from the main channel, Store bit strings, and Make computation on bit strings, with unlimited computation and storage power. But when $\xi$ desires to infer a private information generated by $A$'s DRG (or by $B$'s DRG) from public information, he can only do it in respect of the Deep Random assumption $(A)$. This assumption creates a 'virtual' side channel for the observer conditioning the optimal information $Z$ he can obtain to estimate the Secret Common Output information $X$.

This assumption is fully reasonable, as established in the former sections, under the condition that the DRG of $A$ and the DRG of $B$ can actually produce distributions that are truly indistinguishable and unpredictable among a set $R_G$.

The protocols that we considered in [5] obey the Kerckhoffs's principle by the fact that their specifications are entirely public. We can thus modelize the usage of such protocol in 2 phases:

|  | Legitimate partners | Opponent |
| --- | --- | --- |
| The elaboration phase | The specification of the protocol is made public<br><br>($\mathcal{P}$ with notations of section below) | The observer elaborate its strategy, being a deterministic or probabilistic function taking as parameters the public information that are released during an instantiation<br><br>($Z(\cdot,\cdot)$ with notations of section below) |
| The instantiation phase | The legitimate partners both compute their estimation of the Secret Common Output information based on (i) their part of the secret information generated during an instantiation, and (ii) the public information that is released during the instantiation<br><br>(respectively $X(u,i,j)$ for $A$ and $Y(v,i,j)$ for $B$ with notations of section below) | The Opponent computes its estimation of the Secret Common Output information as the value taken by its strategy function with the released public information as input parameters.<br><br>($Z(i,j)$ with notations of section below) |

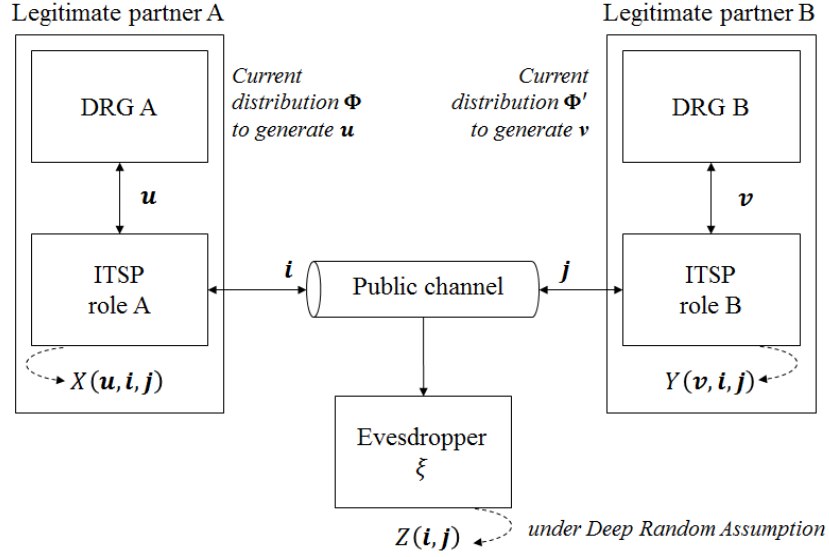

In this model, $A$ (resp. $B$) requests its Deep Random Generator (DRG) to obtain an experiment $u$ (resp. $v$) of a private random variable with hidden probability distribution $\Phi$ (resp. $\Phi'$); $A$ (resp. $B$) publishes the set of information $i$ (resp. $j$) on the public channel along the protocol. $A$ calculates the Secret Common Output information $X(u,i,j)$, with value in metric space $F$. $B$ calculates its estimation $Y(v,i,j)$ of the Secret Common Output information, also with value in $F$. In this model, $\Im_<$ is the public information available regarding the DRG of $A$ or $B$ (that are supposed to have the same design), and $\Im_> = \{i,j\}$ is the set of information published by the partners along the execution of the protocol.

The DRG of $A$ is run by $A$ completely privately and independently of $B$, and reversely the DRG of $B$ is run by $B$ completely privately and independently of $A$. The 2 DRG are thus not in any way secretly correlated, as one of the assumptions of the security model is that $A$ and $B$ share initially no private information.

The eavesdropping opponent $\xi$ who has a full access to the public information, calculates its own estimation $Z(\Im)$ (that we will also shortly denote $Z(i,j)$). $Z$ is called 'strategy' of the opponent.

As introduced in the 'Deep Random assumption' section, $\Im_<$ designates the public information available about a DRG. We assume here that $\Im_<$ is the same for both DRG of $A$ and $B$, meaning that they have the same design.

From the Deep Random assumption, for any group $G$ in $\Gamma_\Im$, the set of optimal strategies for the opponent can be restricted to:

$$\Omega_\Im(G,\mathcal{P}) = \{Z(\Im) = E[X|\Im] | \forall (\varphi_\Im(u), \varphi'_\Im(v)) \in R_\Im(G) \times R_\Im(G)\}$$

Then, the protocol $\mathcal{P}$ is an Information Theoretically Secure Protocol (ITSP) if it verifies the following property $(P)$:

$$\exists G \in \Gamma_\Im, \alpha > 0 | \forall Z \in \Omega_\Im(G,\mathcal{P}): H(X|Z) - H(X|Y) \geq \alpha \quad (P)$$

It has been shown ([7], [8]) that when a protocol satisfies $H(X|Z) - H(X|Y) \geq \alpha$, it can be complemented with Reconciliation and Privacy Amplification techniques to reach Unconditional Security.

We will also impose a second condition for the definition of an ITSP $\mathcal{P}$. The second condition ensures that one can implement a DRG suitable for the protocol $\mathcal{P}$ thanks to a recursive and continuous generation algorithm that emulates locally the protocol. This approach is presented in section IV and introduced below. The second condition is the following:

$\exists \boldsymbol{\alpha} > \boldsymbol{0}$ such that for any strategy $\boldsymbol{Z(\mathfrak{I}) \in \{E[X|\mathfrak{I}] | \forall \psi(u) \in D_{\mathfrak{I}_<}\}}$, there exists an actual distribution $\boldsymbol{\Psi(u) \in D_{\mathfrak{I}_<}}$ of the variables $X$ and $Y$ that verifies $\boldsymbol{H(X|Z) - H(X|Y) \geq \alpha}$

$$(\boldsymbol{P'})$$

The condition $(P')$ means that for any optimal strategy $Z(\mathfrak{I})$ for the distribution $\psi$, there exists a new distribution $\Psi(u) \in D_{\mathfrak{I}_<}$, such that the condition $H(X|Z) - H(X|Y) \geq \alpha$ is satisfied (we take $P_u = P_v = \Psi$ to calculate $H(X|Y)$ because symmetry in the roles $A$ and $B$ can be assumed when their DRG have same design). $\Psi$ does not depend on $\mathfrak{I}_>$ because the generation of a distribution by a DRG takes place before any instantiation of the protocol. The interest of that condition is that (i) it is not referring to the Deep Random Assumption (as it does not involve $\mathfrak{I}_>$), and therefore (ii) it enables to build a Deep Random Generator as follows: at each step $m + 1$, the generator emulates the ITSP internally and picks (through classical randomness) a new couple of identical distributions $\Phi_{m+1}(u) \in D_{\mathfrak{I}_<}, \Phi_{m+1}'(v) = \Phi_{m+1}(v)$ that defeats the optimal strategy (thus belonging to $\{E[X|\mathfrak{I}] | \forall \psi(u) \in D_{\mathfrak{I}_<}\}$) for the past distributions for $t \leq m$. This is always possible for an ITSP as given by condition $(P')$. We will use this recursive method in the following to design a DRG algorithm. The source of secret entropy is the current values of the inifinite counters (several can run in parallel) of the continuous recursive process, together with the classical random that is used at each step to pick a defeating distribution. Such entropy must of course be at least equal to $\log|\mathcal{X}|$ in order to approach perfect secrecy as close as desired.

## II. Notations and first considerations

We will consider in the following a Deep Random generation process that needs to output, when requested, a parameter vector in $[0,1]^n$.

Let's now introduce our formalism and define some notations to ease the use of general Bernoulli random variables ; considering $x = (x_1, \ldots, x_n)$ and $y = (y_1, \ldots, y_n)$ some parameter vectors in $[0,1]^n$ and $i = (i_1, \ldots, i_n)$ and $j = (j_1, \ldots, j_n)$ some experiment vectors in $\{0,1\}^n$, $l, r \in \mathbb{N}_n^*$ two integers, and $\theta \in [0,1]$, we define:

$$x.y \text{ (resp. } i.j\text{) the scalar product of } x \text{ and } y \text{ (resp. } i \text{ and } j\text{)}$$

$$|x| \triangleq \sum_{s=1}^n x_s \; ; |i| \triangleq \sum_{s=1}^n i_s$$

$$i \subset j \Leftrightarrow \{s | i_s = 1\} \subset \{s | j_s = 1\}$$

$$\bar{i} \triangleq (u_1, \ldots, u_n) \mid u_s = 1 \Leftrightarrow i_s = 0$$

$$i \backslash j \triangleq (u_1, \ldots, u_n) \mid u_s = 1 \Leftrightarrow (i_s = 1) \& (j_s = 0)$$

$$i \cap j \triangleq (u_1, \ldots, u_n) \mid u_s = 1 \Leftrightarrow (i_s = 1) \& (j_s = 1)$$

$$i \cup j \triangleq (u_1, \ldots, u_n) \mid u_s = 1 \Leftrightarrow (i_s = 1) \text{ or } (j_s = 1)$$

$$\chi_i(x) \triangleq \prod_{s=1}^{n}(i_s x_s + (1-i_s)(1-x_s)) = P(i|x)$$

that represents probability of obtaining $i$ in a draw of a Bernoulli random variable with parameter vector $x$.

$$\Pi_i(x) \triangleq \prod_{s=1}^{n}(i_s x_s + (1-i_s))$$

$$\psi_{i,r}(x) \triangleq \sum_{j|j.i=r} \chi_j(x) \; ; \; \psi_l(x) = \psi_{(1,\ldots,1),l}(x) \triangleq \sum_{j||j|=l} \chi_j(x)$$

$$\beta_{l,r}(\theta) \triangleq \binom{r}{l}\theta^l(1-\theta)^{r-l}$$

that represents the Bernoulli coefficient of parameter $\theta$.

We will also manipulate permutation operators over vectors. For $\sigma \in \mathfrak{S}_n$, we write $supp(\sigma) = \ker(\sigma - id_{\mathfrak{S}_n}) = \{i, \sigma(i) \neq i\}$ and $|\sigma| = card(supp(\sigma))$. The permutation of a vector is the following linear application :

$$\forall \sigma \in \mathfrak{S}_n, \sigma(x) \triangleq (x_{\sigma(1)}, \ldots, x_{\sigma(n)}) \text{ where } \mathfrak{S}_n \text{ represents the symmetric group}$$

The computations about Bernoulli random variable take place in the vector space of multinomials with $n$ variables and local degree 1, noted $\mathcal{M}_1(n)$, whose $\{\Pi_i(x)\}_{i \in \{0,1\}^n}$ and $\{\chi_i(x)\}_{i \in \{0,1\}^n}$ are basis. It is easy to justify that $\{\chi_i(x)\}_{i \in \{0,1\}^n}$ is a basis by noting the inversion formula:

$$\sum_{i \in \{0,1\}^n} \chi_i(x) = 1 \Longrightarrow \Pi_j(x) = \sum_{i \supset j} \chi_i(x)$$

To ease the manipulation of upper bounds, we will use the notation $\triangleleft (M)$ for any quantity with value in $\mathbb{R}$ and whose absolute value is bounded by $M$.

If $A$ and $B$ are two entities, owning private parameter vectors $x/k$ and $y/k$ respectively (where $k$ is a real number parameter $> 1$), and generating vectors $i$ and $j$ respectively through Bernoulli random variables from $x/k$ and $y/k$ respectively, we consider the respective random variables $V_A$ and $V_B$ such that:

$$V_A = \frac{x.j}{n} \; ; \; V_B = \frac{i.y}{n}$$

By direct computation:

$$E[V_A] = E[V_B] = \frac{x.y}{nk}$$

$$Var(V_A)^2 = E[(V_A - E[V_A])^2] = \frac{1}{nk}\left(\frac{1}{n}\sum_{r=1}^{n} x_r^2 y_r \left(1 - \frac{y_r}{k}\right)\right)$$

$$Var(V_B)^2 = E[(V_B - E[V_B])^2] = \frac{1}{nk}\left(\frac{1}{n}\sum_{r=1}^{n} y_r^2 x_r \left(1 - \frac{x_r}{k}\right)\right)$$

$$E[(V_B - V_A)^2] = \frac{1}{n^2 k}\sum_{r=1}^{n} x_r y_r \left(x_r + y_r - \frac{2 x_r y_r}{k}\right) \leq \frac{2}{nk}$$

The observer $\xi$ who has access only to $(i,j)$ and who wants to engender a random variable with same expected value than $V_A$ can do no more than estimate it from $(i,j)$. The only possible variable to get a strict equality of expectation is:

$$E[V_\xi | i, j] = \frac{k i . j}{n}$$

∎ As $V_A$ takes value in $[0,1]$, the engendered variable has necessarily a distribution function shaped as :

$$P(V_\xi = t | i, j) = v_{i,j}(t) \chi_i\left(\frac{x}{k}\right) \chi_j\left(\frac{y}{k}\right)$$

Then :

$$E[V_\xi] = \int_0^1 \sum_{i,j \in \{0,1\}^n} t\, v_{i,j}(t) \chi_i\left(\frac{x}{k}\right) \chi_j\left(\frac{y}{k}\right) dt = E[V_A] = \frac{x.y}{nk}$$

Considering that $\{\chi_i(x)\chi_j(y)\}_{i,j \in \{0,1\}^n}$ is a basis for the vectorial space $(\mathcal{M}_1(n))^2$ and that :

$$\sum_{i,j \in \{0,1\}^n} \frac{k i . j}{n} \chi_i\left(\frac{x}{k}\right) \chi_j\left(\frac{y}{k}\right) = \frac{x.y}{nk}$$

we deduce that $\int_0^1 t v_{i,j}(t) dt = \frac{k i . j}{n}$ ∎

On the other hand, we observe that the distance from $V_\xi$ to $V_A$, called the evaluation gap between $\xi$ and $A$, by applying Schwarz inequality can be made greater than the variance of $V_A$ when $k \to \infty$

$$E\left[(V_\xi - E[V_A])^2\right] \geq E\left[\left(\frac{k i . j}{n} - E[V_A]\right)^2\right] = \frac{1}{n}\left(\frac{1}{n}\sum_{r=1}^{n} x_r y_r \left(1 - \frac{x_r y_r}{k^2}\right)\right) \geq k\, E[(V_A - E[V_A])^2]$$

which means that if $\xi$ wants to evaluate $V_A$ from the public information $(i,j)$ with exactly the same expectation, it is necessarily less accurate than the legitimate partners.

However, the observer has no obligation to estimate $V_A$ with a variable having exactly the same expectation than $V_A$. As soon as the observer knows the probability distribution that enables $A$ and $B$ to independently generate respectively $x$ and $y$, it can estimate from the public information $(i,j)$, by using bayesian inference, a random variable $\omega_{i,j}$ such that:

$$E[(\omega - E[V_A])^2] \leq E[(V_A - E[V_A])^2]$$

$E[(\omega - E[V_A])^2]$ can be expressed by:

$$E[(\omega - E[V_A])^2] = \int_{x,y \in [0,1]^n} \sum_{i,j \in \{0,1\}^n} \left(\omega_{i,j} - \frac{x.y}{nk}\right)^2 \chi_i\left(\frac{x}{k}\right) \chi_j\left(\frac{y}{k}\right) \Phi(x)\Phi'(y) dx dy$$

where $\Phi$ and $\Phi'$ are called the probability distribution of $x$ and $y$, and $\{\omega_{i,j}\}_{i,j \in \{0,1\}^n}$ is called the Strategy of the observer. In the above example, the optimal strategy is expressed by:

$$\omega^*_{i,j} = \frac{\int_{x,y \in [0,1]^n} \frac{x.y}{nk} \chi_i\left(\frac{x}{k}\right) \chi_j\left(\frac{y}{k}\right) \Phi(x)\Phi'(y) dx dy}{\int_{x,y \in [0,1]^n} \chi_i\left(\frac{x}{k}\right) \chi_j\left(\frac{y}{k}\right) \Phi(x)\Phi'(y) dx dy} = E\left[\frac{x.y}{nk} | i,j\right]$$

that clearly depends on the knowledge of $\Phi$ and $\Phi'$.

We can obtain easily obtain the following upper-bounds, showing the efficiency of the optimal strategy:

**Proposition 0.**

*We have :*

(i) $\inf_\omega E\left[(\omega_{i,j} - V_B)^2\right] \leq E[(V_A - V_B)^2]$.

(ii) $\inf_\omega E\left[(\omega_{i,j} - E[V_B])^2\right] \leq \frac{6}{nk}$

■ For (i), we write $E[(V_A - V_B)^2]$ and simply apply Schwarz inequality :

$$E[(V_A - V_B)^2] = \int_{x,y \in [0,1]^n} \sum_{i,j \in \{0,1\}^n} \left(\frac{x.j}{n} - \frac{i.y}{n}\right)^2 \chi_i\left(\frac{x}{k}\right) \chi_j\left(\frac{y}{k}\right) \Phi(x)\Phi'(y) dx dy$$

$$= \int_y \sum_{i,j} \left[\int_x \left(\frac{x.j}{n} - \frac{i.y}{n}\right)^2 \chi_i\left(\frac{x}{k}\right) \Phi(x) dx\right] \chi_j\left(\frac{y}{k}\right) \Phi'(y) dy$$

$$\geq \int_y \sum_{i,j} \left(\int_x \chi_i\left(\frac{x}{k}\right) \Phi(x) dx\right) \left(\frac{\int_x \frac{x.j}{n} \chi_i\left(\frac{x}{k}\right) \Phi(x) dx}{\int_x \chi_i\left(\frac{x}{k}\right) \Phi(x) dx} - \frac{i.y}{n}\right)^2 \chi_j\left(\frac{y}{k}\right) \Phi'(y) dy$$

By remarking that $\frac{\int_x \frac{x.j}{n} \chi_i\left(\frac{x}{k}\right) \Phi(x) dx}{\int_x \chi_i\left(\frac{x}{k}\right) \Phi(x) dx}$ is of the form $\omega(i,j)$ the result follows. Let's denote $\omega^*_B$ a strategy satisfying $\inf_\omega E\left[(\omega_{i,j} - V_B)^2\right]$.

For (ii):

$$\inf_\omega E\left[(\omega_{i,j} - E[V_B])^2\right] \leq E[(\omega^*_B - V_B + V_B - E[V_B])^2] \leq 2(E[(\omega^*_B - V_B)^2] + E[(V_B - E[V_B])^2])$$
$$\leq \frac{6}{nk}$$

from previous computation. ■

The strategy of the observer is obtained by quadratic optimization (conditional expectation), we simply observe by Schwarz inequality that a strategy is actually a deterministic function of the public information, taking values within the same sample space than the evaluated random variable. This justifies the notation $\omega_{i,j} \in [0, 1/k]$.

In all what follows, the observer strategy will then always be a deterministic function of the public information.

When Bernoulli random variables are considered through quadratic evaluation, the probability distribution $\Phi$ of a legitimate partner can be considered through its quadratic matrix:

$$M_\Phi(u, v) = \int_{[0,1]^n} x_u x_v \Phi(x) dx$$

$M_\Phi$ is a symmetric matrix with elements in [0,1].

$n$ is assumed even. The notation $S_n$ represents the set of all subsets $I$ of $\{1, \ldots, n\}$ containing $n/2$ elements. We remind that for $I \in S_n$, the notation $\bar{I}$ designates $\{1, \ldots, n\} \setminus I$ thus also in $S_n$. Let's first introduce the canonical mid-segment:

$$I_0 \triangleq \{1, \ldots, n/2\}$$

and the canonical mid-segment permutation $\sigma_0$ that sends $I_0$ in $\bar{I}_0$ and vice versa:

$$\sigma_0 \triangleq \circ_{r \in I_0} \tau(r, r + n/2)$$

With this, we now introduce the « tidied form » of a probability distribution $\Phi$. For any $\Phi$, there exists a (not necessarily unique) pair of permutations $(\sigma_\Phi^-, \sigma_\Phi^+)$ in $\mathfrak{S}_n$ such that:

$$c_-(\Phi) \triangleq \sum_{u,v \in I_0 \times \bar{I}_0} M_{\Phi \circ \sigma_\Phi^-}(u, v) = \min_{\sigma \in \mathfrak{S}_n} \left( \frac{4}{n^2} \sum_{u,v \in I_0 \times \bar{I}_0} M_{\Phi \circ \sigma}(u, v) \right)$$

$$c_+(\Phi) \triangleq \sum_{u,v \in I_0 \times \bar{I}_0} M_{\Phi \circ \sigma_\Phi^+}(u, v) = \max_{\sigma \in \mathfrak{S}_n} \left( \frac{4}{n^2} \sum_{u,v \in I_0 \times \bar{I}_0} M_{\Phi \circ \sigma}(u, v) \right)$$

We remark that composing $\Phi$ by $\sigma_0$ does not change a pair $(\sigma_\Phi^-, \sigma_\Phi^+)$ due to the symmetry of $M_\Phi$. Actually, we will call $\sigma_\Phi$ a tidying permutation of $\Phi$ as being either (with 50% chance) a $\sigma_\Phi^-$ or a $\sigma_\Phi^+$. In other words, if we designate by $\mathfrak{S}_I$ the sub-group $\mathfrak{S}_I \triangleq \{\sigma \in \mathfrak{S}_n | \forall u \in I, \sigma(u) \in I\}$, then we mean that picking a tidying permutation of $\Phi$ is equivalent to choose a permutation uniformly within $\{\sigma_\Phi^+ \circ \sigma, \sigma_\Phi^+ \circ \sigma \circ \sigma_0, \sigma_\Phi^- \circ \mu, \sigma_\Phi^- \circ \mu \circ \sigma_0 | \forall \sigma, \mu \in \mathfrak{S}_{I_0}\}$ (obviously this set does not depend on which possible pair $(\sigma_\Phi^-, \sigma_\Phi^+)$ is chosen).

We then call a tidied form of $\Phi$, denoted by $\Phi \circ \sigma_\Phi$, a distribution obtained by composing $\Phi$ with any possible tidying permutation picked randomly. In other words, a tidied form of a distribution is a distribution of the form:

$$\tilde{\Sigma}(\Phi) \triangleq \frac{1}{4}\left(\delta_{\sigma_\Phi^+}(\sigma_\Phi) + \delta_{\sigma_\Phi^-}(\sigma_\Phi)\right)\left(\delta_{\sigma_0}(\rho) + \delta_{\mathrm{Id}_{\mathfrak{S}_n}}(\rho)\right)\frac{1}{|\mathfrak{S}_{I_0}|}\sum_{\sigma \in \mathfrak{S}_{I_0}} \Phi \circ \sigma_\Phi \circ \sigma^{-1} \circ \rho^{-1}$$

Then we define the following probability distribution set:

$$\zeta(\alpha) \triangleq \left\{\Phi \,\big|\, |c_+(\Phi) - c_-(\Phi)| \geq \sqrt{\alpha}\right\}$$

### III. Deep Random generation for Bernoulli experiment vectors

In the present Section, we discuss methods to generate Deep Random from a computing source. It may appear difficult to generate Deep Random from a deterministic computable program. In the real world, even if a computer may access sources of randomness whose probability distribution is at least partly unknown, it doesn't mean that we can use it to build reliable Deep Randomness for cryptographic applications. A specialized strategy is needed.

**Generation process and strategies**

We present a theoretically valid method to generate Deep Random from a computing source. That method relies on a recursive and continuously executing algorithm that generates at each step a new probability distribution, based on a Cantor' style diagonal constructing process.

Taking back the notations of Section II, we will denote for more simplicity:

$$\langle \omega, \Phi \rangle \triangleq E\left[\left(\omega_{i,j} - \frac{x.y}{nk}\right)^2\right] = \int_{x,y \in [0,1]^n} \sum_{i,j \in \{0,1\}^n} \left(\omega_{i,j} - \frac{x.y}{nk}\right)^2 \chi_i\left(\frac{x}{k}\right)\chi_j\left(\frac{y}{k}\right)\Phi(x)\Phi(y)dxdy$$

where both $x$ and $y$ are generated with the same probability distribution $\Phi$.

**Proposition 1.**

*Let $\Phi$ be distribution in $\zeta(\alpha)$, then there exists a constant $C$ such that:*

(i) *For any strategy $\omega$, $\frac{1}{n!}\sum_{\sigma \in \mathfrak{S}_n}\langle \omega, \Phi \circ \sigma \rangle \geq \frac{C}{n}$*

(ii) *For any strategy $\omega$, there exists $\sigma_\omega \in \mathfrak{S}_n$ such that $\langle \omega, \Phi \circ \sigma_\omega \rangle \geq \frac{C}{n}$*

∎

*(i)* is an immediate consequence of Lemma 3 of [5], that provides the last inequality below:

$$\frac{1}{n!}\sum_{\sigma \in \mathfrak{S}_n}\langle \{\omega_{i,j}\}, \Phi \circ \sigma \rangle = \frac{1}{n!}\sum_{\sigma \in \mathfrak{S}_n}\langle \{\omega_{\sigma(i),\sigma(j)}\}, \Phi \rangle \geq \langle\left\{\frac{1}{n!}\sum_{\sigma \in \mathfrak{S}_n}\omega_{\sigma(i),\sigma(j)}\right\}, \Phi \rangle \geq \frac{C}{n}$$

the first inequality being obtained by using Schwarz inequality. And *(ii)* is an immediate consequence of *(i)* as the average summing is lesser than the maximum element.

∎

Let's present a heuristic argument. With a recursive algorithm privately and continuously executed by a partner, the steps $m$ and $m + 1$ are indistinguishable for the observer $\xi$. If a set $\Omega_m$ of winning strategies at the moment of step $m$ exists for $\xi$, then for any probability distribution $\Phi$, by using Schwarz inequality:

$$\frac{1}{|\Omega_m|} \sum_{\omega \in \Omega_m} \langle \omega, \Phi \rangle \geq \langle \frac{1}{|\Omega_m|} \sum_{\omega \in \Omega_m} \omega, \Phi \rangle$$

and thus, by choosing at step $m + 1$ the probability distribution $\Phi_{m+1}$ such that:

$$\langle \frac{1}{|\Omega_m|} \sum_{\omega \in \Omega_m} \omega, \Phi_{m+1} \rangle \geq \frac{C}{n} \quad \left( \gg_{k \gg 1} E[(V_B - V_A)^2] = O\left(\frac{1}{nk}\right) \right)$$

(which is always possible as stated by Proposition 1) the partner guarantees that no absolute winning strategy exists, because the current step at the moment of observation cannot be determined by observer as rather being $m$ or $m + 1$.

This heuristic argument does not explain how to practically build a Deep Random generator with classical computing resources, but it introduces the diagonal constructing process inspired from non-collaborative games' theory. Of course, no Nash' style equilibrium can be found as the partner is continuously deviating its distribution strategy to avoid forecast from observer.

Another interesting remark at this stage is that, in Proposition 1, we highlighted low entropy subsets of probability distributions (typically $\{\Phi \circ \sigma | \forall \sigma \in \mathfrak{S}_n\}$) in which a distribution can always be found to fool a given strategy $\omega$. Such subsets have entropy of order $O(n \ln n)$ although the whole set of possible probability distributions has an entropy of order $O(2^n)$, which is not manageable by classical computing resources.

Let's now express a recursive algorithm, inspired from the above heuristic argument.

Let's first consider the two following recursive algorithms:

Algorithm 1:

$\Phi_0$ is a distribution in $\zeta(\alpha)$ ; at step $m$ :

    i)    $\widehat{\omega}_m$ is performing a minimum value of $\min_\omega \langle \omega, \Phi_m \rangle$
    ii)   $\Psi_{m+1}$ is chosen in $\zeta(\alpha)$
    iii)  $\sigma_{m+1}$ in chosen such that $\langle \widehat{\omega}_m, \Psi_{m+1} \circ \sigma_{m+1} \rangle \geq \frac{C}{n}$ which is always possible as per Proposition 1
    iv)  $\Phi_{m+1} = \Psi_{m+1} \circ \sigma_{m+1}$

Algorithm 2:

$\Phi_0$ is a distribution in $\zeta(\alpha)$ ; at step $m$ :

    i)    $\widehat{\omega}_m$ is performing a minimum value of $\min_\omega \langle \omega, \frac{1}{m} \sum_{s=1}^m \Phi_s \rangle$
    ii)   $\Psi_{m+1}$ is chosen in $\zeta(\alpha)$

iii) $\sigma_{m+1}$ in chosen such that $\langle \hat{\omega}_m, \Psi_{m+1} \circ \sigma_{m+1}\rangle \geq \frac{C}{n}$ which is always possible as per Proposition 1

iv) $\Phi_{m+1} = \Psi_{m+1} \circ \sigma_{m+1}$

The algorithm 1 involves fast variations, but, if the choices of $\Psi_{m+1}$ and $\sigma_{m+1}$ are deterministic at each step $m$, it can have short period (typically period of 2), which makes it unsecure. The algorithm 2 has no period but its variations are slowing down when $m$ is increasing. We can justify that the algorithm 2 has no period by seeing that, if it would have a period, then $\frac{1}{m}\sum_{s=1}^{m} \Phi_s$ would converge, which would contradict the fact that $\langle \hat{\omega}_m, \Phi_{m+1}\rangle \geq \frac{C}{n}$.

Thus, by combining the two algorithms, we get a sequence with both fast variations and no periodic behavior. A method to combine both algorithms is the following:

The Recursive Generation algorithm:

$\Phi_0$ *is a distribution in* $\zeta(\alpha)$ ; *at step* $m$ :

i) $\hat{\omega}_m$ *is performing a minimum value of* $\min_\omega \langle \omega, \Phi_m\rangle$

ii) $\hat{\omega}'_m$ *is performing a minimum value of* $\min_\omega \langle \omega, \frac{1}{m}\sum_{s=1}^{m} \Phi_s\rangle$

iii) $\Psi_{m+1}$ *and* $\Psi'_{m+1}$ *are chosen randomly in* $\zeta(\alpha)$

iv) $\sigma_{m+1}$ *is chosen such that* $\langle \hat{\omega}_m, \Psi_{m+1} \circ \sigma_{m+1}\rangle \geq \frac{C}{n}$ *which is always possible as per Proposition 1*

v) $\sigma'_{m+1}$ *is chosen such that* $\langle \hat{\omega}'_m, \Psi'_{m+1} \circ \sigma'_{m+1}\rangle \geq \frac{C}{n}$ *which is always possible as per Proposition 1*

vi) $\Phi_{m+1} = \alpha_{m+1}\Psi_{m+1} \circ \sigma_{m+1} + (1-\alpha_{m+1})\Psi'_{m+1} \circ \sigma'_{m+1}$ *where* $\alpha_{m+1} \in [0,1]$ *can be picked randomly*

Such an algorithm is secure « against the past », even if the choices at each step are deterministic, but it is not secure « against the future » if the choices at each step are deterministic. Being secure « against the future » means that if the observer runs the recursive algorithm on its own and is « in advance » compared to the legitimate partner, it still cannot obtain knowledge about the current value of $\Phi_m$. So, for the algorithm to remain secure also against the future, it is necessary that the choices at each step involve classical randomness with maximum possible entropy among $\zeta(\alpha)$.

It is also important that the DRG is capable to pick at each step a new distribution within the widest possible sub-set of $\zeta(\alpha)$, otherwise, any restriction in such possible 'picking set' would add prior information for the opponent about the possible distribution, and would therefore reduce $\Im_< \supseteq \zeta(\alpha)$.

To that regards, it can be noted that, in the above algorithm, the distribution $\Psi_{m+1}$ (resp. $\Psi'_{m+1}$) can actually be chosen tidied in $I_0$ (e.g. $\sigma_{\Psi_{m+1}}(I_0) = \sigma_{\Psi'_{m+1}}(I_0) = I_0$) because anyway it is further recomposed by a permutation $\sigma_{m+1}$ (resp. $\sigma'_{m+1}$). From there, we also remark that one can reduce the entropy of choosing a new distribution tidied in $I_0$ by observing that, for any such distribution $\Phi$ and for any $\sigma, \mu \in S_{I_0} \times S_{\bar{I_0}}$, a partner has no reason to choose $\Phi$ rather than $\Phi \circ \sigma \circ \mu$, and therefore any such distribution can be considered of the form:

$$F_{DRG} \triangleq \left\{ \frac{1}{((n/2)!)^2} \sum_{\substack{\sigma \in S_{I_0} \\ \mu \in S_{\overline{I_0}}}} \Phi \circ \sigma \circ \mu \mid \Phi \in \zeta(\alpha) \right\}$$

Then, by denoting

$$L_\Phi(r,s) \triangleq \int_{\substack{|x_{|I_0}|=r \\ |x_{|\overline{I_0}}|=s}} \Phi(x) dx$$

it is clear with a bit of attention that:

$$\frac{1}{((n/2)!)^2} \sum_{\substack{\sigma \in S_{I_0} \\ \mu \in S_{\overline{I_0}}}} \Phi \circ \sigma \circ \mu(x) = \frac{2^n}{\binom{n/2}{|x_{|I_0}|}\binom{n/2}{|x_{|\overline{I_0}}|}} L_\Phi(|x_{|I_0}|, |x_{|\overline{I_0}}|)$$

and thus $F_{DRG}$ is isomorph to a subset of the 2 dimensional set:

$$\left\{ f: \mathbb{N}_{n/2} \mapsto [0,1] \,\middle|\, \sum_{r,s \in \mathbb{N}_{n/2}} f(r,s) = 1 \right\}$$

The exact subset is conditioned by the constraint $\Phi \in \zeta(\alpha)$. Let's express this exact subset. By Lemma 1, we know that, for $\Phi \in \zeta(\alpha)$

$$\int_{\{0,1\}^n} \left( \frac{|x||y|}{n^2} - \frac{x \cdot y}{n} \right)^2 \Phi(x)\Phi(y) dx dy \geq C(\alpha) + O\left(\frac{1}{n}\right)$$

Now, if $\Phi$ also belongs to $F_{DRG}$ we can directly calculate the left term and obtain:

$$\int_{\{0,1\}^n} \left( \frac{|x||y|}{n^2} - \frac{x \cdot y}{n} \right)^2 \Phi(x)\Phi(y) dx dy = \left( \sum_{r,s \in \mathbb{N}_{n/2}} \frac{(r-s)^2}{n^2} L_\Phi(r,s) \right)^2 + O\left(\frac{1}{n}\right)$$

Finally, we obtain the simpler expression of $F_{DRG}$ that clearly represents $F_{DRG}$ a 2-dimensional set with reduced entropy of $O(n^2)$:

$$F_{DRG} = \left\{ \Phi(x) = \frac{2^n}{\binom{n/2}{|x_{|I_0}|}\binom{n/2}{|x_{|\overline{I_0}}|}} L_\Phi(|x_{|I_0}|, |x_{|\overline{I_0}}|) \mid L_\Phi: \mathbb{N}_{n/2} \mapsto [0,1], \sum_{r,s \in \mathbb{N}_{n/2}} L_\Phi(r,s) \right.$$

$$\left. = 1, \sum_{r,s \in \mathbb{N}_{n/2}} \frac{(r-s)^2}{n^2} L_\Phi(r,s) \geq \sqrt{C(\alpha)} + O\left(\frac{1}{n}\right) \right\}$$

Choosing the weights of $L_\Phi(r,s)$ can be implemented by using a congruential generator in order to avoid storing the full $[0,1]^{(n/2)^2}$ structure, but it is then recommended that the entropy of the seed of the generator is at least of the size of the entropy of what is published about each new (used) distribution $\Phi$, e.g. $H(\{i, \sigma_\Phi\}) \leq n(1 + \log n)$.

**Argument about the minimum number of steps to reach the maturity period of the Recursive Generation Process**

We consider a convex and compact subset $\Omega$ of possible strategies for the observer (for any subset, its closed convex envelop corresponds to that characteristic). We consider in the followings the sequence of probability distributions $\{\Phi_m\}_{m \in \mathbb{N}^*}$ constructed as follows :

$$\Phi_1 = \Phi$$

where $\Phi$ is a sleeked distribution in $\zeta(\alpha)$, called the seed of the sequence ; $\widehat{\omega}_m$ is performing a minimum value in :

$$\min_{\omega \in \Omega} \frac{1}{m} \sum_{s=1}^{m} \langle \omega, \Phi_s \rangle$$

$\Phi_{m+1}$ is chosen such that :

$$\langle \widehat{\omega}_m, \Phi_{m+1} \rangle \geq \frac{C}{n}$$

**Diagonal Sequence Property.**

*There exists two constants $C'$ and $C''$ such that, for any $N$ verifying $\frac{N}{\ln(N)} \geq C' \dim(\Omega)$,*

$$\min_{\omega \in \Omega} \frac{1}{N} \sum_{s=1}^{N} \langle \omega, \Phi_s \rangle \geq \frac{C''}{n}$$

∎

More precisely, we chose $\Phi_{m+1}$ as:

$$\Phi_{m+1} = \frac{1}{2}(1 + \Phi \circ \sigma_m) \qquad (DSP.1)$$

where $\sigma_m$ is a permutation such that $\langle \widehat{\omega}_m, \Phi \circ \sigma_m \rangle \geq \frac{C}{n}$.

Let's set:

$$\widehat{Q}_m(\omega - \widehat{\omega}_m) \triangleq \frac{1}{m} \sum_{s=1}^{m} \langle \omega, \Phi_m \rangle - \frac{1}{m} \sum_{s=1}^{m} \langle \widehat{\omega}_m, \Phi_m \rangle$$

$$\widehat{\Sigma}_m \triangleq \frac{1}{m} \sum_{s=1}^{m} \langle \widehat{\omega}_m, \Phi_s \rangle$$

$$\langle \omega, \Phi_m \rangle = Q_m(\omega - \omega_m) + \theta_m$$

$\hat{Q}_m$ and $Q_m$ are positive quadratic forms over $\omega \in \Omega$ derived by respective distributions $\frac{1}{m}\sum_{s=1}^{m} \Phi_s(x)\Phi_s(y)$ and $\Phi_m(x)\Phi_m(y)$. $(DSP.1)$ ensures that $\hat{Q}_m$ and $Q_m$ are non-degenerate in $\Omega$. By considering a basis that is simultaneously $\hat{Q}_m$-orthonormal and $Q_m$-orthogonal, we obtain the 2 following relations (remind that $\hat{\omega}_m$ is the minimum over the quadratic form $\hat{Q}_m$):

$$\frac{1}{m}\sum_{s=1}^{m}\langle\omega,\Phi_s\rangle - \frac{1}{m}\sum_{s=1}^{m}\langle\hat{\omega}_m,\Phi_s\rangle = \|\omega - \hat{\omega}_m\|_{\hat{Q}_m}^2$$

$$|\langle\omega,\Phi_{m+1}\rangle - \langle\hat{\omega}_m,\Phi_{m+1}\rangle| \le 2\sqrt{\dim(\Omega)}\|\omega - \hat{\omega}_m\|_{\hat{Q}_m}$$

We have then:

$$\frac{1}{m+1}\sum_{s=1}^{m+1}\langle\omega,\Phi_s\rangle = \frac{m}{m+1}\left(\hat{\Sigma}_m + \left(\frac{1}{m}\sum_{s=1}^{m}\langle\omega,\Phi_s\rangle - \frac{1}{m}\sum_{s=1}^{m}\langle\hat{\omega}_m,\Phi_s\rangle\right)\right)$$

$$+ \frac{1}{m+1}(\langle\omega,\Phi_{m+1}\rangle - \langle\hat{\omega}_m,\Phi_{m+1}\rangle) + \frac{1}{m+1}\langle\hat{\omega}_m,\Phi_{m+1}\rangle \qquad (DSP.2)$$

Applying the 2 above relations to $(DSP.2)$, we get:

$$\hat{\Sigma}_{m+1} \ge \frac{m}{m+1}\hat{\Sigma}_m + \frac{m}{m+1}\|\omega - \hat{\omega}_m\|_{\tilde{\Omega}}^2 - \frac{2\sqrt{\dim(\Omega)}}{m+1}\|\omega - \hat{\omega}_m\|_{\tilde{\Omega}} + \frac{1}{m+1}\frac{C}{4n}$$

$$\ge \frac{m}{m+1}\hat{\Sigma}_m + \frac{1}{m+1}\frac{C}{4n} - \frac{\dim(\Omega)}{m(m+1)}$$

where the last inequality is obtained by taking the minimum of $\frac{m}{m+1}X^2 - \frac{2\sqrt{\dim(\Omega)}}{m+1}X$ over $X$.

Let's consider the recurring sequence:

$$(m+1)X_{m+1} = mX_m + \frac{C}{4n} - \frac{\dim(\Omega)}{m}$$

$$X_1 = \hat{\Sigma}_1$$

By recurring argument, $\hat{\Sigma}_m \ge X_m$ and thus :

$$\hat{\Sigma}_m \ge \frac{C}{4n} - \dim(\Omega)\, O\left(\frac{\ln(m)}{m}\right)$$

which achieves the proof. ∎

If we consider first $\Omega$ as the full set of strategies (e.g. $\left[0,\frac{1}{k}\right]^{2^{2n}}$), the dimension is $2^{2n}$ and thus the constructing algorithm must be iterated exponential times to reach positive lower bound. On the other hand, we make the excessive assumption that the observer has a full knowledge of the distributions chosen by the partner at each step $m$ (of the sequences $\{\Phi_s\}_{s\le m}$) to be able to build optimal $\hat{\omega}_m$. In practice, the observer has never more knowledge about a distribution than a draw of 1 degraded vector $i$, which means that we can restrict the set of strategies to a subset $\Omega^*$ with dimension $O(n)$, and thus the recommended number of iterations for a recursive constructing process before being able to trustfully pick a distribution is $O(n \ln(n))$.

With the same reasoning than the one of Diagonal Sequence Property, we can obtain a more general result:

**Generalized Diagonal Sequence Property.**

*There exists two constants $C'(\alpha)$ and $C''(\alpha)$ such that, for any $N$ verifying $\frac{N}{\ln(N)} \geq C' \dim(\Omega)$, and for any $\{\Phi_1, ..., \Phi_N\} \in \zeta(\alpha)$ and $\{\Phi'_1, ..., \Phi'_N\} \in \zeta(\alpha)$, and any permutation $\sigma_s$ synchronizing $\Phi_s$ and $\Phi'_s$ :*

$$\min_{\omega \in \Omega} \frac{1}{N} \sum_{s=1}^{N} \langle \omega, \Phi_s, \Phi'_s \circ \sigma_s \rangle \geq \frac{C''}{n}$$

This result also means that when $\{\Phi_1, ..., \Phi_N\}$ and $\{\Phi'_1, ..., \Phi'_N\}$ are generated by DRGs, if an observer choose the same strategy $\omega$ at each step $s$, then whatever is that strategy, in average we have $\langle \omega, \Phi_s, \Phi'_s \circ \sigma_s \rangle \gg_{k \gg 1} E[(V_A - V_B)^2]$ as soon as $\frac{N}{\ln(N)} \geq C' \dim(\Omega)$, and the observer therefore loose its capacity to estimate $V_A$ as efficiently as $B$. In other words the distributions become unknown for the observer as soon as $\frac{N}{\ln(N)} \geq C' \dim(\Omega)$, otherwise the observer would be able estimate $V_A$ as efficiently as $B$ by Bayesian inference.

**Algorithmic implementation**

In order to be able to execute the recursive process described in the previous sections, it is necessary to be able to implement en efficient computing routine to determine, given 2 distributions $\Phi_s$ and $\Phi'$, a permutation $\sigma_{s+1}$ such that ($\widehat{\omega}_s$ being the strategy performing a minimum value of $\min_\omega \langle \omega, \Phi_s \rangle$):

$$\langle \widehat{\omega}_s, \Phi' \circ \sigma_{s+1} \rangle \geq \frac{C}{n}$$

It is the main objective of this technical article to present a workable algorithm to do so. It is generally not an easy problem to determine $\sigma_{s+1}$ because, while the optimal strategy $\omega_{i,j} = E[V_A | i, j]$ can easily be expressed mathematically, it cannot be practically computed. We will propose 2 algorithmic methods.

**First algorithmic method: the bilinear approximation**

The first proposed algorithm relies on a bilinear approximation of the observer strategy. General strategies $\omega_{i,j}$ are members of the set $[0,1]^{2^{2n}}$. We consider the bilinear approximation strategies of the form:

$$V_\Omega(i,j) = i \cdot k\Omega j$$

where $\Omega$ is a matrix in $\mathcal{M}_n(\mathbb{R})$. With formerly defined notation, and by also denoting,

$$\mu_u = \int_{[0,1]^n} x_u \Phi(x) dx$$

one can then write the expression of $E[(V_\Omega - E[V_A])^2]$ by direct computation:

$$\begin{aligned} E[(V_\Omega - E[V_A])^2] \\ &= \frac{1}{k} \sum_{u,v,v' \in \mathbb{N}_n^*} \Omega_{u,v} \Omega_{u,v'} \mu_u M_\Phi(v,v') + \frac{1}{k} \sum_{u,u',v \in \mathbb{N}_n^*} \Omega_{u,v} \Omega_{u',v} \mu_v M_\Phi(u,u') \\ &+ \sum_{u,v \in \mathbb{N}_n^*} \Omega_{u,v}{}^2 \mu_u \mu_v + \frac{1}{k^2} \sum_{u,v,u',v' \in \mathbb{N}_n^*} \Omega_{u,v} \Omega_{u',v'} M_\Phi(u,u') M_\Phi(v,v') \\ &- \frac{2}{nk^2} \sum_{u,v,l \in \mathbb{N}_n^*} \Omega_{u,v} M_\Phi(u,l) M_\Phi(v,l) + \frac{1}{n^2 k^2} \sum_{l,l' \in \mathbb{N}_n^*} M_\Phi(l,l')^2 \end{aligned}$$

The above expression can be seen as a bilinear form over the vector space $\mathcal{M}_n(\mathbb{R})$, whose matrix can be written:

$$\mathcal{Q}_\Phi = \left( \left( \mu_u \delta_{u,u'} + \frac{1}{k} M_\Phi(u,u') \right) \left( \mu_v \delta_{v,v'} + \frac{1}{k} M_\Phi(v,v') \right) \right)_{(u,u'),(v,v')}$$

and its quadratic equation being:

$$E[(V_\Omega - E[V_A])^2] = Q(\Omega) = \Omega \cdot \mathcal{Q}_\Phi \Omega - \frac{2}{nk^2} M_\Phi{}^2 \cdot \Omega + \frac{tr(M_\Phi{}^2)}{n^2 k^2}$$

**Proposition 2:**

*There exists a constant $C'''$ such that:*

$$\text{Inf}_{\Omega \in \mathcal{M}_n(\mathbb{R})} E[(V_\Omega - E[V_A])^2] \leq \frac{C'''}{nk}$$

*and $C''' \leq 5$.*

■ We set:

$$D_\Phi \triangleq \begin{pmatrix} \mu_1 & & 0 \\ & \ddots & \\ 0 & & \mu_n \end{pmatrix}, \quad \widetilde{M}_\Phi \triangleq D_\Phi + \frac{1}{k} M_\Phi$$

By considering $\mathcal{Q}_\Phi$ as a block matrix, we easily determine its inverse:

$$\mathcal{Q}_\Phi{}^{-1} = \left( \cdots \left( \widetilde{M}_\Phi{}^{-1} \right)_{u,v} \widetilde{M}_\Phi{}^{-1} \cdots \right)$$

The minimization of $Q(\Omega)$ thanks to $\mathcal{Q}_\Phi{}^{-1}$ is well known, and we can then compute the value of the minimum depending on $D_\Phi$ and $\widetilde{M}_\Phi$:

$$\inf_{\Omega \in \mathcal{M}_n(\mathbb{R})} Q(\Omega) = -\frac{1}{4}\left(-\frac{2}{nk^2}M_\Phi{}^2\right) \cdot Q_\Phi{}^{-1}\left(-\frac{2}{nk^2}M_\Phi{}^2\right) + \frac{tr(M_\Phi{}^2)}{n^2 k^2}$$

$$= \frac{1}{n^2}\Big(2tr(\widetilde{M}_\Phi D_\Phi) + 4tr\left(D_\Phi{}^3 \widetilde{M}_\Phi{}^{-1}\right) - 3tr(D_\Phi{}^2) - 2tr\left(D_\Phi \widetilde{M}_\Phi D_\Phi \widetilde{M}_\Phi{}^{-1}\right)$$

$$- tr\left(\left(D_\Phi{}^2 \widetilde{M}_\Phi{}^{-1}\right)^2\right)\Big) \tag{p2.1}$$

Let's set:

$$L \triangleq D_\Phi{}^{-\frac{1}{2}} M_\Phi D_\Phi{}^{-\frac{1}{2}}$$

$L$ enables to write:

$$\widetilde{M}_\Phi = D_\Phi + \frac{1}{k} M_\Phi = D_\Phi{}^{-\frac{1}{2}}\left(I_n + \frac{1}{k}L\right) D_\Phi{}^{-\frac{1}{2}}$$

$L$ is symmetric positive (and even positive-definite if $\Phi$ is not a Dirac distribution), because:

$$\sum_{u,v \in \{1,\dots,n\}} L(u,v) X_u X_v = \int_{[0,1]^n} \left(\sum_{r \in \{1,\dots,n\}} \frac{x_r}{\sqrt{\mu_r}} X_r\right)^2 \Phi(x) dx \geq 0$$

Therefore all eigenvalues are positive and we can write the diagonalization of $L$ with unitary matrix $U$ (we denote $\widetilde{L}$ the diagonal reduction of $L$):

$$L = U \begin{pmatrix} \tilde{\lambda}_1 & & 0 \\ & \ddots & \\ 0 & & \tilde{\lambda}_n \end{pmatrix} {}^t U \triangleq U \widetilde{L} \, {}^t U$$

By replacing $\widetilde{M}_\Phi$ by

$$\widetilde{M}_\Phi = D_\Phi{}^{-\frac{1}{2}} U \left(I_n + \frac{1}{k}\widetilde{L}\right) {}^t U D_\Phi{}^{-\frac{1}{2}}$$

in all the terms under the trace operator in the right hand term of $(p2.1)$, and by reminding that the elements of the unitary matrix $U$ obey:

$$\forall u,v \quad \sum_{r=1}^n |U_{r,u} U_{r,v}| \leq \left(\sum_{r=1}^n |U_{r,u}|^2 \sum_{r=1}^n |U_{r,v}|^2\right)^{1/2} \leq 1$$

we can manage to obtain the following bounds:

$$\left|tr(\widetilde{M}_\Phi D_\Phi) - tr(D_\Phi{}^2)\right| \leq \frac{tr(L)}{k}$$

$$\left|tr\left(D_\Phi{}^3 \widetilde{M}_\Phi{}^{-1}\right) - tr(D_\Phi{}^2)\right| \leq \frac{tr(L)}{k}$$

$$\left|tr\left(D_\Phi \widetilde{M}_\Phi D_\Phi \widetilde{M}_\Phi{}^{-1}\right) - tr(D_\Phi{}^2)\right| \leq \frac{tr(L)}{k}$$

$$\left| tr\left(\left(D_\Phi{}^2 \widetilde{M}_\Phi{}^{-1}\right)^2\right) - tr\left(D_\Phi{}^2\right) \right| \leq \frac{2tr(L)}{k}$$

And thus eventually we get:

$$\inf_{\Omega \in \mathcal{M}_n(\mathbb{R})} Q(\Omega) \leq \frac{5tr(L)}{n^2 k} \leq \frac{5}{n^2 k} \sum_{u=1}^{n} \frac{\int_{[0,1]^n} x_u{}^2 \Phi(x)dx}{\int_{[0,1]^n} x_u \Phi(x)dx} \leq \frac{5}{nk} \qquad \blacksquare$$

We will denote in the following $\Omega_\Phi$ the bilinear form that realizes $\text{Inf}_{\Omega \in \mathcal{M}_n(\mathbb{R})} E[(V_\Omega - E[V_A])^2]$, and $V_{\Omega_\Phi}$ the associated bilinear strategy.

The bilinear approximation $\Omega_\Phi$ is much easier to manipulate as its size is $O(n^2)$ instead of $O(2^{2n})$. In particular, it is much easier to determine $\sigma_{s+1}$ with $V_{\Omega_{\Phi_s}}$ than with $\widehat{\omega}_s$. We will admit without proof that:

$$\exists M, M' | \forall \Phi_s, \Phi' \in \zeta(\alpha), \quad \langle V_{\Omega_{\Phi_s}}, \Phi' \circ \sigma_{s+1} \rangle \geq M \implies \langle \widehat{\omega}_s, \Phi' \circ \sigma_{s+1} \rangle \geq M' \qquad (cj1)$$

This conjecture is certainly not easy to prove in the general case, but it is verified in simulations. The proposition 2 gives an indication that the approximation is efficient as it gives the same order of magnitude $O\left(\frac{1}{nk}\right)$ than for the optimal strategy $\widehat{\omega}_s$.

Then, determining $\sigma_{s+1}$ such that $\langle V_{\Omega_{\Phi_s}}, \Phi' \circ \sigma_{s+1} \rangle \geq M$ is equivalent to determine $\sigma_{s+1}$ verifying:

$$\frac{1}{n^2} \int_{x,y \in [0,1]^n} \left(x \cdot \frac{1}{k}\Omega_{\Phi_s} y - x \cdot y\right)^2 \Phi' \circ \sigma_{s+1}(x) \Phi' \circ \sigma_{s+1}(y) dx dy \geq M - O\left(\frac{1}{n}\right)$$

The above can be computed by exploring:

$$\min_{\sigma \in \mathfrak{S}_n} \left( \sum_{r,s,u,v=1}^{n} \left(\frac{1}{k}\Omega_{\Phi_s} - I_n\right)_{\sigma(r),\sigma(s)} \left(\frac{1}{k}\Omega_{\Phi_s} - I_n\right)_{\sigma(u),\sigma(v)} M_{\Phi'}(r,u) M_{\Phi'}(s,v) \right)$$

or

$$\max_{\sigma \in \mathfrak{S}_n} \left( \sum_{r,s,u,v=1}^{n} \left(\frac{1}{k}\Omega_{\Phi_s} - I_n\right)_{\sigma(r),\sigma(s)} \left(\frac{1}{k}\Omega_{\Phi_s} - I_n\right)_{\sigma(u),\sigma(v)} M_{\Phi'}(r,u) M_{\Phi'}(s,v) \right)$$

Those explorations can be performed by finding recursively a transposition that decreases (resp increases) the above expression. Such method does not guarantee to find the optimum, but it enables to exceed a bound $M$ with a good probability of success; it follows preparatory sums that are of complexity $O(n^3)$; the search of a decreasing (resp. increasing) permutation is then of complexity $O(n^2)$, and one can show that the number of steps before it stops decreasing (resp. increasing) is $O(n)$. Before starting such exploration, it is of course needed to determine the matrix $M_{\Phi'}$ and $\Omega_{\Phi_s}$, whose main cost is the computation of $\widetilde{M}_{\Phi_s}{}^{-1}$. This inverse can be obtained by the recursive method

$$X_{q+1} - \lambda I_n = X_q\left(I_n - \lambda \widetilde{M}_{\Phi_s}\right)$$

for $\lambda$ chosen strictly inferior to the spectral radius of $\widetilde{M}_{\Phi_s}$. By noting that $\widetilde{M}_{\Phi_s}$ is a positive matrix, the algorithm converges with exponential speed. Its complexity, by using Strassen algorithm for each matrix product, is $(n^{\log 7} \log n)$ to obtain a residual error in $O(1/n)$. Thus, the overall complexity of the exploration process is $O(n^3)$.

**Second algorithmic method: the partition summing**

This method is based on the restriction of the form of the optimal strategy obtained by considering the symmetries within the generator.

For $I, i \in \{0,1\}^n$, we denote $\mathfrak{S}_{I,i}$ is the sub-group of $\mathfrak{S}_n$ that let stable $i \cap I$, $i \cap \bar{I}$, $\bar{i} \cap I$ and $\bar{i} \cap \bar{I}$: $\mathfrak{S}_{I,i} \triangleq \{\sigma \in \mathfrak{S}_n | \forall \{u,v\} \in I \times i, \{\sigma(u), \sigma(v)\} \in I \times i\}$. We remind that the notation $\bar{I}$ designates the complement of $I$ in $\{0,1\}^n$, and the notation $|I|$ designates the cardinality of $I$. If $\Phi$ is a hidden distribution generated by $A$'s Deep Random Generator, and if the observer only knows $\mu$, the assumed value of a synchronization permutation $\sigma_\Phi$ of $\Phi$ and $i$ issued from a Bernoulli trial of parameter vector $x$ generated by $\Phi$, all happen for the observer as if $A$ would perform the sequence below:

$$\Phi \circ \sigma_\Phi : x'' \xrightarrow{\sigma \in \mathfrak{S}_{I_0, \mu^{-1}(i)}} x' \xrightarrow{\mu} x$$

At the first step $x''$ is generated by a certain distribution $\Phi \circ \sigma_\Phi$ (synchronized on $I_0$), at second step $x' = \sigma^{-1}(x'')$ corresponds to a mixing simultaneously within $I_0 \cap \mu^{-1}(i)$, $I_0 \cap \overline{\mu^{-1}(i)}$, $\bar{I}_0 \cap \mu^{-1}(i)$, and $\bar{I}_0 \cap \overline{\mu^{-1}(i)}$ and at third step, $\mu$ produces the final $x$. The resulting distribution to be considered is denoted:

$$\Sigma_i(\Phi, \mu) \triangleq \frac{1}{|\mathfrak{S}_{I_0, \mu^{-1}(i)}|} \sum_{\sigma \in \mathfrak{S}_{I_0, \mu^{-1}(i)}} \Phi \circ \sigma_\Phi \circ \sigma \circ \mu^{-1} = \frac{1}{|\mathfrak{S}_{\mu(I_0), i}|} \sum_{\sigma \in \mathfrak{S}_{\mu(I_0), i}} \Phi \circ \sigma_\Phi \circ \mu^{-1} \circ \sigma$$

In the following, $x|_I$ designates the restriction of the vector $x$ to its components having their index within $I$.

It has been shown in [5] (Proposition 9) the following result:

**Proposition 3.**

*$\mu$ represents any permutation independent of $i, j$. We have the following restricted form for the optimal strategy:*

$$E\left[\frac{\mu^{-1}(x) \cdot \mu'^{-1}(y)}{nk} \bigg| i, j\right]_{\Sigma_i(\Phi, \mu), \Sigma_j(\Phi', \mu')}$$
$$= \frac{2k}{n^2}\left((\mu^{-1}(i) \cdot I_0)(\mu'^{-1}(j) \cdot I_0) + (\mu^{-1}(i) \cdot \bar{I}_0)(\mu'^{-1}(j) \cdot \bar{I}_0)\right) + O\left(\frac{1}{k^2}\right)$$

In the recursive generation process, at step $m$, the optimal strategy can therefore be approximated by:

$$\widehat{\omega}_m(i,j) = \frac{2k}{n^2}\left((\sigma_m^{-1}(i) \cdot I_0)(\sigma_m'^{-1}(j) \cdot I_0) + (\sigma_m^{-1}(i) \cdot \bar{I}_0)(\sigma_m'^{-1}(j) \cdot \bar{I}_0)\right)$$

It becomes then easier to search a sub-set $J$ with $|J| = n/2$, such that, for a given new distribution $\Psi \in \zeta(\alpha)$ tidied in $J$ (e.g. $\sigma_\Psi^{-1}(I_0) = J$), we obtain:

$$k^2 \langle \hat{\omega}_m, \tilde{\Sigma}(\Psi) \circ \sigma_\Psi^{-1} \rangle \geq M$$

Indeed

$$k^2 \langle \hat{\omega}_m, \tilde{\Sigma}(\Psi) \circ \sigma_\Psi^{-1} \rangle$$
$$= \int_{x,y \in [0,1]^n} \left( \frac{2}{n^2} \left( (x \cdot \sigma_m(I_0))(y \cdot \sigma_m(I_0)) + (x \cdot \overline{\sigma_m(I_0)})(y \cdot \overline{\sigma_m(I_0)}) \right) \right.$$
$$\left. - \frac{x \cdot y}{n} \right)^2 \tilde{\Sigma}(\Psi) \circ \sigma_\Psi^{-1}(x) \tilde{\Sigma}(\Psi) \circ \sigma_\Psi^{-1}(y) dx dy + O\left( \frac{1}{n} \right)$$

The distribution $\tilde{\Sigma}(\Psi)$ has a quadratic matrix of the form:

$$M_{\tilde{\Sigma}(\Psi)} = \begin{pmatrix} \alpha & \cdots & \beta & \gamma & \cdots & \gamma \\ \vdots & \ddots & \vdots & \vdots & & \vdots \\ \beta & \cdots & \alpha & \gamma & \cdots & \gamma \\ \gamma & \cdots & \gamma & \bar{\alpha} & \cdots & \bar{\beta} \\ \vdots & & \vdots & \vdots & \ddots & \vdots \\ \gamma & \cdots & \gamma & \bar{\beta} & \cdots & \bar{\alpha} \end{pmatrix}$$

Thus,

$$\int_{x,y \in [0,1]^n} \left( \frac{2}{n^2} \left( (x \cdot \sigma_m(I_0))(y \cdot \sigma_m(I_0)) + (x \cdot \overline{\sigma_m(I_0)})(y \cdot \overline{\sigma_m(I_0)}) \right) - \frac{x \cdot y}{n} \right)^2 \tilde{\Sigma}(\Psi) \circ \sigma_\Psi^{-1}(x) \tilde{\Sigma}(\Psi)$$
$$\circ \sigma_\Psi^{-1}(y) dx dy$$

can be computed as a function of $\alpha_\Psi, \beta_\Psi, \gamma_\Psi$, and $J \cdot \sigma_m(I_0)$, which enables to find easily a value of $J \cdot \sigma_m(I_0)$ maximizing $\langle \hat{\omega}_m, \tilde{\Sigma}(\Psi) \circ \sigma_\Psi^{-1} \rangle$ with error $O(1/k^2)$. The details of the computation are not developed here because of they are painful without presenting any conceptual difficulty. The complexity of this search is $O(1)$ once $\alpha_\Psi, \beta_\Psi, \gamma_\Psi$ are computed. The computation of $\alpha_\Psi, \beta_\Psi, \gamma_\Psi$ has a complexity of $O(n^2)$. Thus, the overall complexity of the exploration process is $O(n^2)$, which makes the second method more efficient than the first one.

The computation of $\alpha_\Psi, \beta_\Psi, \gamma_\Psi$ involves a step of tidying for the quadratic matrix $M_\Psi$. We give below an efficient tidying algorithm:

*Tidying Algorithm*

    Step 1: compute for $s \in \{1, \ldots, n\}$, the value $U_s = \sum_{u \in I_0} M_\Psi(u,s) - \sum_{u \in \bar{I}_0} M_\Psi(u,s)$

    Step 2: order $\{U_s\}$ so that $\forall r \in I_0, \forall s \in \bar{I}_0$, then $U_r \geq U_s$. An obvious way of doing this to compute the sorting permutation $U_{\sigma(1)} \geq \cdots \geq U_{\sigma(n)}$

    Step 3: then $\sigma$ is a tidying permutation of $\Psi$.

Here is a proof:

**Proposition 4.**

    *The tidying algorithm above ensures that $\sum_{u,v \in I_0 \times \bar{I}_0} M_\Psi(\sigma(u), \sigma(v))$ is minimal*

- First, due to the fact that $c_-(\cdot)$ is a norm (see Proposition 5 below), we can always modify the matrix $M_\Psi$ by infinitesimal perturbation to ensure that all the values of

$$\left\{ \sum_{u,v \in I_0 \times \overline{I_0}} M_\Psi(\mu(u), \mu(v)) \mid \forall \mu \in \mathfrak{S}_n \right\}$$

are distinct. This ensures that the segment $J \in S_n$ realizing $\min_{I \in S_n} \sum_{u,v \in I \times \overline{I}} M_\Psi(u,v)$ is unique.

We remark that, for $r \in I_0, s \in \overline{I_0}$, the transposition $\tau$ permuting $r$ and $s$ has the following property:

$$\sum_{u,v \in I_0 \times \overline{I_0}} \left( M_\Psi(\tau(u), \tau(v)) - M_\Psi(u,v) \right) = U_r - U_s - \left( M_\Psi(s,s) + M_\Psi(r,r) - 2M_\Psi(r,s) \right)$$

We have first

$$M_\Psi(s,s) + M_\Psi(r,r) - 2M_\Psi(r,s) = \int_{[0,1]^n} (x_s^2 + x_r^2 - 2x_r x_s) \Psi(x) dx \geq 0$$

Therefore, let's suppose that $\sum_{u,v \in I \times \overline{I}} M_\Psi(u,v)$ is already minimal, then if its vector $\{U_s\}$ is not ordered, then there would exist $r \in I_0, s \in \overline{I_0}$ such that $U_r \leq U_s$, and thus:

$$\sum_{u,v \in I_0 \times \overline{I_0}} \left( M_\Psi(\tau(u), \tau(v)) - M_\Psi(u,v) \right) < 0$$

which would contradict minimality. This means that the vector is necessarily ordered. By unicity, we then deduce reversely that if the vector is ordered, then $\sum_{u,v \in I \times \overline{I}} M_\Psi(u,v)$ is minimal. ∎

The complexity of step 1 is $O(n^2)$, the step 2 is a classical sorting procedure whose complexity is $O(n \log n)$. Then the overall complexity does not exceeed $O(n^2)$ as claimed.

**Proposition 5.**

$c_-(\cdot)$ is a norm on the subspace of the matrix without their main diagonal, for any $n > 4$.

■ Let's justify by showing that $c_-(\Phi) = 0 \Leftrightarrow M_\Phi(u,v) = 0 \ \forall u \neq v$:

$\Leftarrow$ is obvious.

For $\Rightarrow$, we calculate $\frac{n^4}{16}\sum_{I\in S_n}\left(\sum_{u,v\in I\times \bar{I}} M_\Phi(u,v)\right)^2$:

First we have:

$$\sum_{\substack{I\in S_n \\ I\times\bar{I} \ni (u,v),(u',v')}} 1 = \binom{n-2}{\frac{n}{2}-2}$$

$$\sum_{\substack{I\in S_n \\ I\times\bar{I}\ni(u,v),(u,v')}} 1 = \binom{n-2}{\frac{n}{2}-1}$$

$$\sum_{\substack{I\in S_n \\ I\times\bar{I}\ni(u,v)}} 1 = \binom{n-1}{\frac{n}{2}-1}$$

$$\frac{n^4}{16}\sum_{I\in S_n} c_I(M)^2 = \sum_{I\in S_n}\sum_{(u,v),(u',v')\in I\times\bar{I}} M(u,v)M(u',v')$$

$$= \sum_{\substack{(u,v),(u',v') \\ u\neq v, u\neq u' \\ u'\neq v', v\neq v'}} \left(\sum_{\substack{I\in S_n \\ I\times\bar{I}\ni(u,v),(u',v')}} 1\right) M(u,v)M(u',v')$$

$$+ 2\sum_{\substack{u,v,v' \\ u\neq v, u\neq v' \\ v\neq v'}}\left(\sum_{\substack{I\in S_n \\ I\times\bar{I}\ni(u,v),(u,v')}} 1\right) M(u,v)M(u,v') + \sum_{\substack{u,v \\ u\neq v}}\left(\sum_{\substack{I\in S_n \\ I\times\bar{I}\ni(u,v)}} 1\right) M(u,v)^2$$

$$= \binom{n-2}{\frac{n}{2}-2}\left(\sum_{\substack{u,v \\ u\neq v}} M(u,v)\right)^2 + 2\left(\binom{n-2}{\frac{n}{2}-1}-\binom{n-2}{\frac{n}{2}-2}\right)\sum_u\left(\sum_{\substack{v \\ v\neq u}} M(u,v)\right)^2$$

$$+ \left(\binom{n-1}{\frac{n}{2}-1}+\binom{n-2}{\frac{n}{2}-2}-2\binom{n-2}{\frac{n}{2}-1}\right)\sum_{\substack{u,v \\ u\neq v}} M(u,v)^2$$

We conclude by noticing that:

$\binom{n-2}{\frac{n}{2}-1} - \binom{n-2}{\frac{n}{2}-2} > 0$ and $\binom{n-1}{\frac{n}{2}-1} + \binom{n-2}{\frac{n}{2}-2} - 2\binom{n-2}{\frac{n}{2}-1} = \binom{n-2}{\frac{n}{2}-1}\frac{n-4}{n} > 0$ for $n > 4$. ∎

## IV. Conclusion

We have given in this article a polynomial time algorithm that can be recursively executed to form a Deep Random generator within a classical computing resource. Such Deep Random generator is the core component of Deep Random secrecy introduced in [5]. It is the first of several articles aiming at proposing a practical implementation of Deep Random Secrecy. Besides Deep Random Secrecy, Deep Random generators may have other applications in contexts where it is desired to prevent third parties to make efficient predictions over a probabilistic model; one could think about preventing speculative trading for instance.

Further work on the ideas and methods introduced in this article could typically cover theoretical aspects, like finding a rigorously proven method to generate $\sigma_{s+1}$ (proving the conjecture ($cj1$), or proving that the general problem can always be lower-bounded by using a partition summing method, are two ways of doing so). It could also cover practical aspects like finding the fastest possible algorithm to compute an acceptable $\sigma_{s+1}$.

Further work on Deep Random generation in general could typically pursue a more general characterization of what Deep Random generators could be, then triggering research for optimal implementation for each category of them. Within the category of recursive generators running on classical computing resources, a set of general and rigorously defined characteristics that they should verify would be welcome, in order to organize the research on this topic. Such characteristics should typically cover the Deep Random assumption introduced in [5], and its application to secrecy systems.

**Who is the author ?**
I have been an engineer in computer science for 20 years. My professional activities in private sector are related to IT Security and digital trust, but have no relation with my personal research activity in the domain of cryptology. If you are interested in the topics introduced in this article, please feel free to establish first contact at tdevalroger@gmail.com

**Annex 1: An impossibility result**

One could be tempted to try to reduce the entropy of the set of distributions coursed by a DRG to a polynomial dimension set. We show here a result stating that it is or impossible or unsecure.

Let's first define what we mean by reducing entropy. We consider $F$ a subset of distributions of $\zeta(\alpha)$. We define the associated set of optimal strategies $U(F)$ by:

$$U(F) \triangleq \left\{ E\left[\frac{x \cdot y}{nk} \Big| i,j\right]_{\Phi,\Phi} \Big| \Phi \in F \right\}$$

Then we say that $F$ is called $C$-stable iff:

$$\forall \omega \in U(F), \quad \exists \Phi \in F: \langle \omega, \Phi \rangle \geq \frac{C}{n}$$

One can understand this definition as the possibility for a DRG to execute the recursive algorithm described above by keeping the picking of each new distribution $\Phi_s$ within $F$.

We have the following result (with the same notations than for the Diagonal Sequence Property):

**Theorem A1:**

*Any subset $F$ of $\zeta(\alpha)$ that is both $C''$-stable and convex has a superpolynomial dimension*

■ By contradiction. Let's assume that the dimension of $F$ can be bounded by $n^q$. As a consequence of the Diagonal Sequence Property, we can extract from $F$ a sequence $\{\Phi_s\}$ such that, for any $N$ verifying $\frac{N}{\ln(N)} \geq C' \dim(F)$:

$$\min_{\omega \in \Omega} \frac{1}{N} \sum_{s=1}^{N} \langle \omega, \Phi_s \rangle \geq \frac{C''}{n} \tag{A1.1}$$

We denote by $\widetilde{\Phi}_N$ the distribution $\widetilde{\Phi}_N \triangleq \frac{1}{N} \sum_{s \in \mathbb{N}_N} \Phi_s$. $\widetilde{\Phi}_N$ is still in $F$ because $F$ is convex, and $(A1.1)$ means that when $x, y$ are results of draws from both the distribution $\widetilde{\Phi}_N$, we can find a polynomial $P(n)$ such that, for any strategy $\omega$, by extracting the square terms in the sum $\widetilde{\Phi}_N(x)\widetilde{\Phi}_N(y)$, we can write (for $\sqrt{k} \gg K' > K \gg 1$):

$$P\left(\left|\omega - \frac{x \cdot y}{nk}\right| \geq \frac{K'}{\sqrt{nk}}\right) \geq \frac{1}{P(n)} \tag{A1.2}$$

On the other hand, we know that there exist a constant $\lambda$ such that (Chernoff style bound; see for instance [5] Proposition 3)

$$P\left(\left|V_A - \frac{x \cdot y}{nk}\right| \geq \frac{K}{\sqrt{nk}}\right) \leq e^{-\lambda K} \tag{A1.3}$$

$$P\left(\left|V_B - \frac{x \cdot y}{nk}\right| \geq \frac{K}{\sqrt{nk}}\right) \leq e^{-\lambda K}$$

$$P\left(|V_A - V_B| \geq \frac{K}{\sqrt{nk}}\right) \leq e^{-\lambda K}$$

Let's now consider the restricted condition $\Gamma_\omega(x,i,j)$, available to the partner $A$ in the emulated protocol:

$$\Gamma_\omega(x,i,j): \quad |\omega(i,j) - V_A(x,j)| \geq \frac{K'}{\sqrt{nk}}$$

For any strategy $\omega$, from $(A1.2)$ and $(A1.3)$ we deduce that one can find have a polynomial $Q(n)$ such that

$$P\left(|V_A - V_B| \geq \frac{K}{\sqrt{nk}} | \Gamma_\omega(x,i,j)\right) \leq e^{-\lambda K} Q(n) \qquad (A1.4)$$

On the other hand, let's consider $G(\omega)$ the optimal strategy realizing:

$$\min_\Omega \left( E[(\Omega(i,j) - V_A)^2 | i,j,\Gamma_\omega(x,i,j)]_{\widetilde{\Phi}_N} \right)$$

$G$ can be written:

$$G(\omega) = E[V_A | i,j,\Gamma_\omega(x,i,j)]_{\widetilde{\Phi}_N} = \frac{\int_{x|\Gamma_\omega(x,i,j)} \frac{x \cdot j}{n} \chi_i\left(\frac{x}{k}\right) \widetilde{\Phi}_N(x) dx}{\int_{x|\Gamma_\omega(x,i,j)} \chi_i\left(\frac{x}{k}\right) \widetilde{\Phi}_N(x) dx}$$

It is a continuous because, due to $(A1.2)$, the measure of $\Gamma_\omega(x,i,j)$ is lower bounded by a strictly positive constant independent of $\omega$, and therefore

$$\forall \omega, \quad \int_{\substack{x,y \\ \Gamma_\omega(x,i,j)}} \chi_i\left(\frac{x}{k}\right) \chi_j\left(\frac{y}{k}\right) \widetilde{\Phi}_N(x), \widetilde{\Phi}_N(y) dx dy > \varepsilon$$

$G$ being both bounded and continuous, has a fix point $\omega^*$ (Brouwer). Then, if we consider the conditional distribution resulting from the discard of all instances not verifying $\Gamma_{\omega^*}(x,i,j)$, it still is a known distribution and therefore, by considering the emulated protocol, we have

$$I(V_B; V_A | i,j) = 0$$

and then Shannon impossibility theorem dictates that the optimal strategy $\omega^*$ verifies the first inequality below; and further more $(A1.4)$ also applies to $\omega^*$ which gives the second inequality below

$$E[(\omega^* - V_A)^2 | i,j,\Gamma_{\omega^*}(x,i,j)]_{\widetilde{\Phi}_N,\widetilde{\Phi}_N} \leq E[(V_B - V_A)^2 | i,j,\Gamma_{\omega^*}(x,i,j)]_{\widetilde{\Phi}_N,\widetilde{\Phi}_N} \leq \frac{K^2}{nk} + e^{-\lambda K} Q(n)$$

This contradicts the fact that, by definition of $\Gamma_{\omega^*}(x,i,j)$

$$E[(\omega^* - V_A)^2 | i,j,\Gamma_{\omega^*}(x,i,j)]_{\widetilde{\Phi}_N,\widetilde{\Phi}_N} \geq \frac{K'^2}{nk}$$

(the quantity $e^{\lambda K}$ is super-polynomial for appropriate choice of $K$ and thus $e^{-\lambda K} Q(n)$ can be made as small as desired, and in particular $< \left(\frac{K'^2}{nk} - \frac{K^2}{nk}\right)$) ∎

This result means that it is impossible to find a convex subset of $\zeta(\alpha)$ of polynomial dimension, where the DRG will be able to always pick a successor in the recursive algorithm.

On the other hand, the subset that we have obtained in the description of the algorithm is not constrained by the pessimistic result of Theorem A1, because, while being of polynomial dimension, it is not convex. Indeed, this set is

$$\{\Phi \circ \sigma | \Phi \in F_{DRG}, \sigma \in \mathfrak{S}_n\}$$

and it is easy to see that while $F_{DRG}$ alone is convex, the composition by $\mathfrak{S}_n$ remove the convexity.